\documentclass[prl, reprint, amsmath, amssymb, floatfix, aps, longbibliography, superscriptaddress]{revtex4-1}
\usepackage{graphicx}
\usepackage{dcolumn}
\usepackage{bm}
\usepackage[colorlinks=true,urlcolor=rossos,anchorcolor=black,citecolor=mygreen,filecolor=black,linkcolor=black,menucolor=blue,linktocpage=true]{hyperref} 
\usepackage{subfigure}

\usepackage{booktabs}
\usepackage{colortbl}
\usepackage{collcell}
\usepackage{enumerate}
\usepackage{float}
\usepackage{verbatim}
\usepackage{multirow}
\usepackage{xparse}
\usepackage{tabularx}
\usepackage{cancel}

\usepackage{multirow}

\usepackage[normalem]{ulem}
\usepackage{soul}
\setstcolor{red}

\definecolor{rossos}{cmyk}{0,1,1,0.55}
\definecolor{mygreen}{rgb}{0.27, 0.64, 0.48}
\definecolor{mygray}{gray}{0.95}

\begin{document}

\title{Probing High-Quality Axions with Gravitational Waves}

\author{Ruiyu Zhou}
%\email{zhoury@cqupt.edu.cn}
\affiliation{School of Science, Chongqing University of Posts and Telecommunications, Chongqing 400065, P.R. China}
\author{ Jin-Wei Wang}
\email{jinwei.wang@uestc.edu.cn}
\affiliation{School of Physics, University of Electronic Science and Technology of China, Chengdu 611731, China}
\author{Ligong Bian}
\email{lgbycl@cqu.edu.cn}
%\thanks{\\ R.Z. and J.-W.W.   contributed equally to this work.\\}
\affiliation{Department of Physics and Chongqing Key Laboratory for Strongly Coupled Physics, Chongqing University, Chongqing 401331, P. R. China}

\begin{abstract}
We present a systematic study of gravitational wave (GW) signals from phase transitions and topological defects in a unified high-quality axion framework. The gauged $U(1)_g$ symmetry forbids any bias term that could lift the vacuum degeneracy, restricting the theory to the phenomenologically viable case $N_{\rm DW}=1$. Requiring the axion to account for the observed dark matter (DM) abundance and satisfy the high-quality condition constrains the gauge symmetry-breaking scale to $f_g \in [1.6\times10^{11},\,10^{16}]\,\mathrm{GeV}$ for the QCD axion, leading to a well-defined band of GW signals, part of which is consistent with current pulsar timing array observations. Two-step first-order phase transitions are common in this framework, with the lower-scale transition generating GWs with $f^{\rm peak} \gtrsim \mathcal{O}(10^7)\,\mathrm{Hz}$. For axion-like realizations, 
%both high-quality quintessence axions and fuzzy DM axions yield no observable GW signals due to inflationary dilution, while 
generic post-inflation models predict GW spectra that are nearly degenerate with the QCD axion case. We conclude that GWs alone cannot distinguish between these scenarios, highlighting the need for complementary probes.
\end{abstract}

\maketitle

\noindent \textit{\textbf{Introduction.}}---It is widely accepted that the Standard Model (SM) represents only an effective theory valid at low energy scales, as it fails to account for several well-established phenomena in our universe, such as dark matter (DM), dark energy (DE), neutrino masses, and the strong-$CP$ problem \cite{hep-ph/0404175,Taoso:2007qk,Bertone:2016nfn,1807.06209,Arbey:2021gdg,Cirelli:2024ssz,Super-Kamiokande:1998kpq,SNO:2002tuh,Peccei:1977hh}. These shortcomings strongly suggest the existence of physics beyond the SM, making the identification of new fundamental degrees of freedom one of the central goals of modern particle physics. In this context, a particularly appealing direction is to construct minimal frameworks that can simultaneously address multiple fundamental problems, in the spirit of Occam’s razor \cite{Profumo:2008ms}.

In this regard, the unified axion framework is especially attractive, as it provides a coherent explanation for DM, DE, and the strong-$CP$ problem within a single setup \cite{Qiu:2023los}. From the perspective of axion model building, two long-standing challenges are the ultraviolet (UV) origin of the axion and the so-called high-quality problem. The axion arises as the Nambu--Goldstone boson of a spontaneously broken global $U(1)$ symmetry, such as the Peccei--Quinn (PQ) symmetry \cite{Wilczek:1977pj,Weinberg:1977ma}, whose fundamental origin remains obscure. Moreover, quantum gravity effects are generally expected to violate global symmetries and induce Planck-suppressed operators that can spoil the axion solution unless they are sufficiently suppressed \cite{Abbott:1989jw,Kallosh:1995hi,Alvey:2020nyh,Ardu:2020qmo}.

Ref.~\cite{Qiu:2023los} provides an elegant resolution to both issues by introducing two complex scalar fields, $\phi_1$ and $\phi_2$, which give rise to two global $U(1)$ symmetries associated with independent phase rotations. A particular linear combination can be promoted to a gauged symmetry, denoted as $U(1)_g$, while the orthogonal combination, $U(1)_a$, remains global and naturally plays the role of the axion symmetry \cite{Fukuda:2017ylt}. The high quality of the axion is then protected by the gauge symmetry $U(1)_g$.

Although this unified high-quality axion framework is compact and theoretically well motivated, its phenomenological implications remain largely unexplored. As a representative paradigm of high-quality axion models, it encompasses a broad class of axion scenarios across hierarchical energy scales, ranging from the QCD axion to axion-like fuzzy DM and quintessence axion. Such high-scale axion sectors are generically inaccessible to conventional laboratory or astrophysical probes. We show that gravitational waves (GWs) provide a unique observational window into this class of models, offering both a powerful discriminator among different realizations of high-quality axion frameworks and a direct probe of new physics at ultra-high energy scales.

In this work, we consider two distinct sources of GWs: those generated by phase transitions (PT) and those arising from the formation of gauged cosmic strings (CS) 
\cite{Kibble:1976sj,Hindmarsh:1994re}. When $\phi_1$ and $\phi_2$ acquire vacuum expectation values (VEVs), both $U(1)_a$ and $U(1)_g$ are spontaneously broken. If these symmetry breakings are first order, they can generate a stochastic gravitational wave background (SGWB) through bubble nucleation, collisions, and plasma-induced turbulence in the early universe \cite{Witten:1984rs,Hogan:1983ixn,Turner:1990rc,Kosowsky:1992vn,Kamionkowski:1993fg,Caprini:2015zlo,Caprini:2019egz,Athron:2023xlk}. In addition, the associated formation of gauged CSs can in principle produce distinct GW signals through their decay.

In the high-quality axion framework of Ref.~\cite{Qiu:2023los}, multiple PTs can occur at hierarchically different energy scales, corresponding to the QCD axion, axion-like fuzzy DM, and a quintessence axion. Each of these transitions, together with the associated topological defects, can leave characteristic imprints in the GW spectrum. These signals encode rich information about the underlying symmetry structure and the UV realization of the high-quality axion mechanism. In this letter, we perform a systematic analysis of GW signatures arising from first-order PTs and topological defects in unified high-quality axion models, and demonstrate that future GW observations may provide a powerful discriminator among different realizations of the high-quality axion framework.\\

\noindent \textit{\textbf{High-quality axion model.}}---The new sector of the model consists of two complex scalar fields, $\phi_1$ and $\phi_2$, together with $N$ pairs of chiral fermions $\{\psi_i,\psi'_i\}$, where $i=1,2,\dots,N$. All new fields are charged under both a global $U(1)_a$ symmetry and a gauged $U(1)_g$ symmetry. The gauge charge assignments are required to satisfy the anomaly cancellation conditions \cite{Qiu:2023los}
\begin{subequations}
\label{eq:anomaly_cancel}
\begin{align}
    \sum_{i=1}^N \!\left(U(1)_g^{\psi_i}+U(1)_g^{\psi'_i}\right) &= 0, \label{eq:anomaly_cancel_1} \\
    \sum_{i=1}^N \!\left[\left(U(1)_g^{\psi_i}\right)^3+\left(U(1)_g^{\psi'_i}\right)^3\right] &= 0,
    \label{eq:anomaly_cancel_2}
\end{align}
\end{subequations}
where $U(1)_g^{\psi_i}$ ($U(1)_g^{\psi'_i}$) denotes the $U(1)_g$ charge of $\psi_i$ ($\psi'_i$).
As emphasized in the context of quantum gravity, ratios of Abelian gauge charges are expected to be rational numbers \cite{Banks:2010zn}. With an appropriate normalization, all $U(1)_g$ charges can therefore be taken to be integers without loss of generality. In addition, the charge assignments must forbid any gauge invariant bare mass terms for the fermions; otherwise, Planck scale masses would be generated, causing these states to decouple from low-energy phenomenology. We thus require that all fermion masses arise solely from Yukawa interactions with the scalar fields. Under this assumption, the $U(1)_g$ charges of the scalars, denoted by $q_{1,2}$, are fixed by gauge invariance of the Yukawa sector.

The detailed charge assignments for both fermions and scalars have been discussed extensively in Ref.~\cite{Qiu:2023los}. In this work, we adopt the same configuration for consistency and do not repeat the derivation here. We assume that $k$ fermion pairs couple to $\phi_1$, while the remaining $l=N-k$ pairs couple to $\phi_2$. The anomaly-cancellation condition in Eq.~\eqref{eq:anomaly_cancel_1} then implies
\begin{equation}
    -\frac{q_1}{q_2}=\frac{l}{k}=\frac{m}{n},
\end{equation}
where $m$ and $n$ are coprime integers. This relation fixes the ratio of scalar charges and plays an important role in determining the effective axion domain wall (DW) number, as discussed below.

The Lagrangian of the new sector is given by
\begin{align}
    \mathcal{L} = \sum_{i=1}^{2}|D_\mu \phi_i|^2
    + \sum_{i=1}^{N} \bar{\psi}_i i\gamma^\mu D_\mu \psi_i
    + \sum_{i=1}^{N} \bar{\psi}'_i i\gamma^\mu D_\mu \psi'_i  \nonumber \\
    - \frac{1}{4}F_{\mu\nu}F^{\mu\nu}
    - \mathcal{L}_{\rm Yukawa}
    - \mathcal{L}_V ,
\end{align}
with
\begin{equation}
    \mathcal{L}_{\rm Yukawa}
    = \sum_{i=1}^k y_i\,\phi_1\,\psi_i\psi'_i
    + \sum_{j=1}^{l} y'_j\,\phi_2\,\psi_j\psi'_j ,
\end{equation}
and the scalar potential
\begin{equation}
    \mathcal{L}_V
    = -m_1^2|\phi_1|^2 - m_2^2|\phi_2|^2
    + \lambda_1|\phi_1|^4 + \lambda_2|\phi_2|^4
    + \kappa|\phi_1|^2|\phi_2|^2 .
    \label{eq:vp}
\end{equation}
Here $y_i$ and $y'_j$ denote the Yukawa couplings, while $\lambda_{1,2}$ and $\kappa$ control the scalar potential. Owing to the $U(1)_g$ gauge symmetry, Eq.~\eqref{eq:vp} already represents the most general renormalizable scalar potential. That is to say,  additional operators, such as $\phi_{1,2}^3$ or $\phi_1\phi_2^3$, are forbidden. This restricted structure significantly enhances the predictivity of the model. 

In our framework, the lowest-dimensional nonrenormalizable operator that preserves the gauged $U(1)_g$ symmetry while explicitly breaking the global $U(1)_a$ symmetry is given by~\cite{Qiu:2023los}
\begin{equation}
    \mathcal{O}
    = \frac{1}{k!\,l!}\,
      \frac{\phi_1^{\,k}\phi_2^{\,l}}{M_{\rm Pl}^{\,N-4}}
      + \text{h.c.},
    \label{eq:PQbreak}
\end{equation}
where $M_{\rm Pl}=2.4\times10^{18}\,\mathrm{GeV}$ denotes the reduced Planck scale. This operator induces a correction to the axion potential of the form~\cite{Qiu:2023los}
\begin{equation}
    \delta V_a
    = \frac{2^{1-N/2}}{k!\,l!}\,
      \frac{f_1^{\,k}f_2^{\,l}}{M_{\rm Pl}^{\,N-4}}
      \cos\!\left(\frac{a\,N_{\rm DW}}{F_a}\right),
      \label{eq:axionV}
\end{equation}
where
\begin{equation}
    F_a=\frac{f_1 f_2}{\sqrt{m^2 f_1^2+n^2 f_2^2}}
    \label{eq:fa}
\end{equation}
is the axion decay constant, and $N_{\rm DW}$ denotes the DW number, given by the greatest common divisor of $(k,l)$. Different levels of axion quality are therefore realized by different choices of $(k,l)$, leading to parametrically distinct suppressions of the explicit $U(1)_a$-breaking effects.\\

\noindent \textit{\textbf{Gravitational wave production.}}---In the unified high-quality axion framework, GWs may arise from three different sources: first-order PTs, CSs, and DWs. In what follows, we discuss these contributions in turn and clarify which of them are relevant in the phenomenologically viable parameter space.

\noindent \textbf{(I) Phase transitions:} 
In the unified high-quality axion framework, the vacuum expectation values of $\phi_1$ and $\phi_2$ spontaneously break both $U(1)_a$ and $U(1)_g$. If these symmetry-breaking transitions are first order, they proceed via the nucleation and expansion of true-vacuum bubbles, sourcing a SGWB. The dominant GW production arises from bubble collisions and the subsequent bulk motion of the plasma in the form of long-lived acoustic waves, while the turbulence contribution remains theoretically uncertain and is therefore neglected \cite{Kosowsky:1992vn,Caprini:2015zlo,Caprini:2019egz}.

A key feature of this setup is the presence of two hierarchical symmetry-breaking scales, the gauge scale $f_g\sim f_2$ and the PQ scale $f_a\sim f_1$. When both transitions are first order, they generically lead to two GW signals with parametrically separated peak frequencies and amplitudes. This multi-scale imprint provides a distinctive observational handle on the high-quality axion framework and opens the possibility of probing ultra-high energy physics through future GW observations. In the following, we focus on the representative case $f_2 \geq f_1$.

To characterize the PT dynamics, we employ the full one-loop finite-temperature effective potential $V_{\rm eff}(\phi_1,\phi_2,T)$, including Coleman--Weinberg corrections, thermal effects, and daisy resummation. The explicit expressions and renormalization conditions are summarized in the Supplemental Material. The bounce solution and the corresponding Euclidean action $S_3$ are obtained using \texttt{FindBounce}~\cite{Guada:2018jek,Guada:2020xnz}.

The GW signal is primarily controlled by two parameters: the strength parameter $\alpha$ and the inverse duration parameter $\beta/H$, defined at the percolation temperature $T_p$ as \cite{Quiros:1999jp,Caprini:2015zlo}
\begin{equation}
    \alpha = \frac{1}{\rho_{\text{rad}}} \left[ \Delta V - T \frac{d\Delta V}{dT} \right]_{T_p},  ~~\frac{\beta}{H} = T \left. \frac{d}{dT} \left( \frac{S_3}{T} \right) \right|_{T_p}.
\end{equation}
Here $\Delta V$ denotes the free-energy difference between the false and true vacua, and $\rho_{\rm rad}$ is the radiation energy density.

The total GW spectrum from PT is given by
\begin{equation}
\Omega_{\rm GW}^{\rm PT}(f) h^2
= \Omega_{\rm coll}(f) h^2
+ \Omega_{\rm sw}(f) h^2,
\label{gw}
\end{equation}
where we include only the bubble-collision and sound-wave contributions. The collision contribution is
\begin{align}
\Omega_{\rm{coll}}(f) h^2\simeq  1.67&\times 10^{-5}\left(\frac{\beta}{H}\right)^{-2}\left(\frac{\kappa_\phi\alpha}{1+\alpha}\right)^2\left(\frac{100}{g^*}\right)^{1/3}\nonumber \\& \times\frac{0.11v_w^3}{0.42+v_w^2}\frac{3.8(f/f_{\rm{coll}})^{2.8}}{1+2.8(f/f_{\rm{coll}})^{3.8}}\;,
\label{eq:co}
\end{align}
while the sound-wave contribution reads~\cite{Hindmarsh:2013xza,Hindmarsh:2015qta,Hindmarsh:2017gnf}
\begin{align}
\Omega_{\rm{sw}}&(f) h^2=2.65\times 10^{-6}(H \tau_{\rm{sw}})\left(\frac{\beta}{H}\right)^{-1}v_w\left(\frac{\kappa_\nu \alpha}{1+\alpha}\right)^2 \nonumber\\&\times\left(\frac{g^*}{100}\right)^{-1/3}\left(\frac{f}{f_{\rm{sw}}}\right)^3\left(\frac{7}{4+3(f/f_{\rm{sw}})^2}\right)^{7/2}\;,
\label{Omega_sw}
\end{align}
The peak frequency is
\begin{equation}
f_{\rm sw} =
1.9\times10^{-5}
\frac{\beta}{H}
\frac{T}{100~{\rm GeV}}
\frac{1}{v_w}
\left(\frac{g_*}{100}\right)^{1/6}
{\rm Hz},
\end{equation}
and the duration of the acoustic period is
\begin{equation}
\tau_{\rm sw} = \min\!\left[\frac{1}{H},\,\frac{R_*}{\bar U_f}\right],
\end{equation}
where $R_*=v_w(8\pi)^{1/3}/\beta$ is the mean bubble separation, $\overline{U}_f^2\approx 3\kappa_\nu \alpha/4(1+\alpha)$ is  the root-mean-square fluid velocity \cite{Hindmarsh:2017gnf,Caprini:2019egz,Ellis:2019oqb}, $\kappa_\nu$ denotes the efficiency factor for the conversion of latent heat into kinetic energy of the fluid $\kappa_\nu=\sqrt{\alpha}/({0.135+\sqrt{0.98+\alpha}})$ \cite{Espinosa:2010hh}.\\

\noindent \textbf{(II) Cosmic strings:} The spontaneous breaking of the gauged $U(1)_g$ symmetry leads to the formation of CSs~\cite{Kibble:1976sj}. The subsequent evolution of the string network is well described by the Nambu--Goto dynamics together with string intercommutation, resulting in a scaling network composed of long strings and loops~\cite{Forster:1974ga,Shellard:1987bv,Blanco-Pillado:2017oxo}. The loops gradually lose energy through gravitational radiation, giving rise to a SGWB.

The present-day GW spectrum from string loops can be expressed as~\cite{Blanco-Pillado:2013qja,Blanco-Pillado:2017oxo}
\begin{align}
    \Omega_{\rm{GW}}^{\rm{CS}}(\ln &f)h^2 = \frac{8\pi \Gamma G^2 \mu^2 f h^2}{3H_0^2 \zeta\left(\frac{4}{3},\infty\right)} \sum_{j=1}^{\infty} j^{-4/3} \frac{2j}{f^2} \nonumber \\
    &\times\int_{0}^{\infty}\frac{dz}{H(z)(1+z)^6} n\left(t(z),\frac{2j}{(1+z)f}\right),
    \label{eq:GWstring}
\end{align}
where $G\mu \sim G\pi f_g^2$ denotes the dimensionless string tension \cite{Hill:1987qx}, $\Gamma\simeq50$ is the loop emission efficiency \cite{Vilenkin:1981bx,Turok:1984cn,Quashnock:1990wv,Blanco-Pillado:2013qja,Blanco-Pillado:2017oxo}, and $\zeta\left(\frac{4}{3},\infty\right) = \sum_{j=1}^{\infty} j^{-4/3}$ arises from the cusp-dominated GW emission spectrum. Here $H(z)$ and $t(z)$ are the Hubble parameter and cosmic time, respectively, and $n\left(t(z),\frac{2j}{(1+z)f}\right)$ is the loop number density. We also include the effect of changes in the relativistic degrees of freedom in the early universe.

For a scaling string network, the loop distribution can be approximated analytically. In the radiation-dominated era, which is relevant for our parameter space, it is given by~\cite{Blanco-Pillado:2013qja,Blanco-Pillado:2017oxo}
\begin{equation}
    n(t,l) = \frac{0.18}{t^{3/2} (l + \Gamma G \mu t)^{5/2}} 
    \, \Theta(0.1 - l/t),
    \label{eq:loopdensity}
\end{equation}
where $l$ denotes the loop length, related to the observed frequency through $l=2j/[(1+z)f]$. Combining Eqs.~\eqref{eq:GWstring} and \eqref{eq:loopdensity}, we compute the GW spectrum from CSs in the unified high-quality axion framework. In this analysis, we adopt the BOS model as a benchmark description of the string network evolution; alternative approaches, such as the VOS and LRS models \cite{Lorenz:2010sm,Cui:2018rwi,Gouttenoire:2019rtn,Gouttenoire:2019kij}, may lead to quantitative differences in the predicted GW spectrum \cite{Luo:2025ewp,Bian:2022tju}.

For the parameter region of interest, the string network decays entirely during the radiation-dominated epoch. The resulting GW spectrum therefore exhibits a nearly scale-invariant plateau at high frequencies, providing a characteristic signature of the $U(1)_g$ breaking scale \cite{Blanco-Pillado:2017oxo}
\begin{equation}
    \Omega_{\rm{GW}}^{\rm{CS}} (\ln f) h^2  = 3.3 \times 10^{-4} \sqrt{\frac{G\pi}{\Gamma}} f_g.
\end{equation}

\noindent \textbf{(III) Domain walls:} 
In addition to CSs, DWs provide another potential source of GWs associated with axion dynamics. As follows from Eq.~\eqref{eq:axionV}, when $N_{\rm DW}\neq 1$ the axion potential exhibits multiple degenerate minima. In this case, different Hubble patches generically relax into distinct vacua after symmetry breaking, and DWs inevitably form at their boundaries~\cite{Kibble:1976sj}.
Such DWs are cosmologically catastrophic. Their energy density scales more slowly than that of both radiation and matter, and thus eventually dominates the total energy density of the universe, leading to overclosure~\cite{Zeldovich:1974uw}. A standard remedy is to introduce a small explicit breaking of the discrete symmetry, which lifts the vacuum degeneracy and generates a pressure difference across the walls, triggering their annihilation. The resulting decay of the DW network releases energy in the form of GWs, contributing to a stochastic background~\cite{Vilenkin:1981zs,Gelmini:1988sf,Coulson:1995nv,Larsson:1996sp,Gleiser:1998na,Hiramatsu:2010yz,Kawasaki:2011vv}.

In the present framework, however, such a resolution is not available. Any bias term capable of lifting the vacuum degeneracy is forbidden by the gauged $U(1)_g$ symmetry (see the discussion below Eq.~\eqref{eq:vp}). In particular, the operator $\delta V_a$ in Eq.~\eqref{eq:PQbreak} already represents the leading symmetry-allowed contribution. Higher-order corrections take the form $(\delta V_a)^n$ with $n\geq 2$ and do not qualitatively alter the degeneracy structure. Consequently, no effective bias is generated, and the DWs remain stable for $N_{\rm DW}\neq 1$, rendering these configurations cosmologically inconsistent.

We therefore restrict our analysis to the case $N_{\rm DW}=1$, which can always be realized in this class of models~\cite{Qiu:2023los}. In this case, DWs are absent, and no GW signal from their decay is expected.\\

\noindent \textit{\textbf{Results.}}---We present our results for two representative realizations of the framework: the QCD axion and axion-like particles (ALPs). We discuss them in turn.

\noindent \textbf{(I) QCD axion case:}
We begin with the QCD axion realization, in which the chiral fermions transform as 
$\psi_i \in (\mathbf{3},1,0)$ and $\psi'_i \in (\mathbf{3^*},1,0)$ under 
$SU(3)_{\rm c} \times SU(2)_{\rm L} \times U(1)_{\rm Y}$. 
To realize a viable high-quality QCD axion, we impose the following requirements: 
(i) the axion accounts for the entire DM abundance; 
(ii) the DW number satisfies $N_{\rm DW}=1$; 
(iii) the $U(1)_g$ breaking scale $f_g$ lies below the inflationary scale $\sim 10^{16}\,\mathrm{GeV}$~\cite{Balkenhol:2025wms}; 
and (iv) the gauge couplings $g_g$ and $g_3$ of $U(1)_g$ and $SU(3)_{\rm c}$ remain perturbative up to the Planck scale, i.e., no Landau poles are encountered.

The QCD axion is subject to the well-known relation between its mass and decay constant~\cite{DiLuzio:2020wdo}
\begin{equation}
    m_a \simeq 5.7\left(\frac{10^{12}~\text{GeV}}{F_a}\right)~\mu\text{eV},
    \label{eq:mafa}
\end{equation}
which effectively reduces the dimensionality of the parameter space. 
The relic abundance is determined by the misalignment mechanism, including contributions from string radiation and the decay of the string–DW network. Recent numerical simulations~\cite{Benabou:2024msj} indicate that the observed DM density is reproduced for 
$m_a \in [40,\,300]~\mu\mathrm{eV}$, corresponding to 
$F_a \in [1.9\times10^{10},\,1.4\times10^{11}]~\mathrm{GeV}$. 
Using Eq.~\eqref{eq:fa} and adopting the convention $f_1 \leq f_2$, one obtains
\begin{equation}
    F_a < f_1 \leq F_a\sqrt{n^2+m^2} \leq f_2.
    \label{eq:f12range}
\end{equation}

Higher-dimensional operators in Eq.~\eqref{eq:PQbreak} shift the minimum of the axion potential and induce a contribution to the effective QCD angle $\bar{\theta}$,
\begin{equation}
    \delta \bar{\theta} = \frac{2^{1-N/2}}{k!\,l!} 
    \frac{f_1^k f_2^l}{M_{\rm Pl}^{N-4}} 
    \frac{1}{F_\pi^2 m_\pi^2},
    \label{eq:theta}
\end{equation}
where $m_\pi$ and $F_\pi$ denote the pion mass and decay constant. 
The high-quality condition requires $\delta \bar{\theta} < 10^{-10}$~\cite{Pospelov:1999mv}. 
Combining Eqs.~\eqref{eq:f12range}–\eqref{eq:theta}, we find that this bound implies a minimum operator dimension $N=12$. 
The corresponding minimal symmetry-breaking scale is achieved for $\{k,l\}=\{5,7\}$, yielding 
$f_g \sim f_2 \simeq 1.6\times10^{11}\,\mathrm{GeV}$. 
Therefore, the allowed range is
\begin{equation}
    f_g \in [1.6\times10^{11},\,10^{16}]~\mathrm{GeV}.
\end{equation}
Within this window, the GW spectrum from CSs can be computed using Eq.~\eqref{eq:GWstring}.

In contrast to the CS contribution, the GW spectrum from first-order PTs depends on the full set of model parameters. 
We focus on the minimal case $N=12$ (e.g., $k=1$, $l=11$), noting that larger $N$ leads to qualitatively similar results. 
Solving Eqs.~\eqref{eq:anomaly_cancel_1}–\eqref{eq:anomaly_cancel_2}, one obtains viable $U(1)_g$ charge assignments. 
A representative solution is
\{$-1.1$, $0.6$, $-0.4$, $0.6$, $-0.4$, $1$, $-0.8$, $0.6$, $-0.4$, $0.9$, $-0.7$, $0.1$\}, 
with $\{q_1,q_2\}=\{2.2,-0.2\}$, where we have rescaled the charges to rational values.
\begin{figure}
    \centering
    \includegraphics[width=0.9\linewidth]{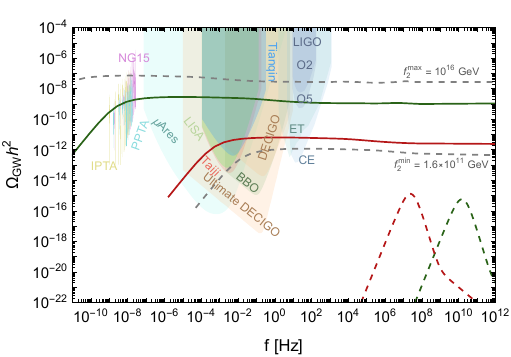}
    \caption{Predicted GW spectra in the high-quality QCD axion framework. The area between the two gray dashed lines represents the range of CS GWs. The colored solid lines and dashed lines represent the GWs of CSs and the PTs, respectively. Violin plots represent PTA measurements from IPTA~\cite{Antoniadis:2022pcn}, PPTA~\cite{Goncharov:2021oub}, and NANOGrav~\cite{NANOGrav:2020bcs}. Shaded regions indicate the sensitivities of current and future detectors, including muAres~\cite{Sesana:2019vho}, LISA~\cite{LISA:2017pwj,Baker:2019nia}, Taiji~\cite{Hu:2017mde,Ruan:2018tsw}, TianQin~\cite{TianQin:2015yph}, DECIGO~\cite{Seto:2001qf,Kudoh:2005as,Kawamura:2020pcg}, BBO~\cite{Crowder:2005nr,Cutler:2005qq} and LIGO-Virgo~\cite{Thrane:2013oya,LIGOScientific:2014qfs,LIGOScientific:2016aoc,LIGOScientific:2019vic}, CE~\cite{Reitze:2019iox}, and ET~\cite{Punturo:2010zz,Hild:2010id,Sathyaprakash:2012jk}.}
    \label{fig:QCDaxionGW}
\end{figure}

These chiral fermions contribute to the renormalization group running of $g_g$ and $g_3$. 
Assuming that new degrees of freedom enter at the scale $f_1$, the requirement of perturbativity up to the Planck scale imposes an upper bound on both the fermion multiplicity and the gauge coupling. 
For $N=12$, we find a maximal value $g_g=1.17$, which we adopt as a benchmark. 
The Yukawa couplings between $\phi_{1,2}$ and $\{\psi_i,\psi'_i\}$ are fixed to $0.5$. 
We scan over the remaining parameters, 
$\{\lambda_1,\lambda_2,\kappa\} \in [0,\pi]$ and 
$\{f_1,f_2\} \in [1.9\times10^{10},\,10^{16}]~\mathrm{GeV}$, 
imposing the constraints in Eq.~\eqref{eq:f12range}, the allowed range of $F_a$, and $\delta \bar{\theta}<10^{-10}$.

We find that a significant fraction of the parameter space exhibits a two-step first-order PT. 
In the following, we focus on the transition associated with $\phi_1$, which typically yields lower-frequency GWs and is therefore more accessible to observation. 
Since only one pair of chiral fermions couples to $\phi_1$ (corresponding to $k=1$), we include only this pair in the effective relativistic degrees of freedom $g_*$ entering Eq.~\eqref{eq:co}.

In Fig.~\ref{fig:QCDaxionGW}, we present the resulting GW spectra from both CSs and PTs. 
The band between two gray dashed lines indicates the predicted range of CS GWs within the high-quality QCD axion framework. 
We highlight two representative parameter points: 
the green curves correspond to a case where the CS signal is compatible with pulsar timing array (PTA) observations, with the dashed curve showing the associated PT contribution; 
the red curves illustrate the lowest-frequency PT signal in our scan and its corresponding CS spectrum. 
For comparison, current PTA data and the projected sensitivities of ongoing and future GW experiments are also shown.\\

\noindent \textbf{(II) Axion-like case:}
We next turn to axion-like realizations. 
As discussed in Ref.~\cite{Qiu:2023los}, both high-quality quintessence axions and fuzzy DM axions can arise in this framework, with chiral fermions transforming as 
$\psi_i \in (1,\mathbf{2},0)$ and $\psi'_i \in (1,\mathbf{2^*},0)$ under 
$SU(3)_{\rm c}\times SU(2)_{\rm L} \times U(1)_{\rm Y}$. 
A crucial feature of these scenarios is that the PQ symmetry is broken before inflation, with $f_{1,2}$ exceeding the inflationary scale. 
As a result, any topological defects formed during the symmetry-breaking PT are inflated away, and the axion field is homogenized over the observable universe. 
Consequently, neither CSs nor PTs contribute to a present-day SGWB.

For general ALPs, the relic abundance in the post-inflation scenario is determined by the misalignment mechanism~\cite{Sheridan:2024vtt},
\begin{equation}
\label{alpomega}
    \Omega_a h^2 \approx 0.12\, \theta_a^2 
\left( \frac{m_a}{4.4 \times 10^{-19}\,\mathrm{eV}} \right)^{1/2}
\left( \frac{F_a}{10^{16}\,\mathrm{GeV}} \right)^2,
\end{equation}
where $\theta_a^2 \simeq \pi^2/3$, and the axion mass is given by
\begin{equation}
    m_a^2=\frac{2^{1-N/2}}{k!\,l!}
    \frac{f_1^k f_2^l}{M_{\rm Pl}^{N-4}}
    \frac{N_{\rm DW}^2}{F_a^2}.
\end{equation}

In contrast to the QCD axion, ALPs in this setup generically couple to the electroweak sector. 
Since the fermions $\{\psi_i,\psi'_i\}$ carry $SU(2)_{\rm L}$ quantum numbers, the $[U(1)_a]\times[SU(2)_{\rm L}]^2$ anomaly is nonvanishing, which induces an effective axion–photon coupling after electroweak symmetry breaking~\cite{Qiu:2023los},
\begin{equation}
    \mathcal{L} \supset \frac{1}{4} g_{a\gamma\gamma} \, a \, F_{\mu\nu} \tilde{F}^{\mu\nu},
\end{equation}
with
\begin{equation}
   g_{a\gamma\gamma} = \frac{g_2^2 \sin^2 \theta_{W}}{8\pi^2} \frac{N_{\rm DW}}{F_a}.
   \label{eq:gagg}
\end{equation}

\begin{figure}
    \centering
    \includegraphics[width=0.9\linewidth]{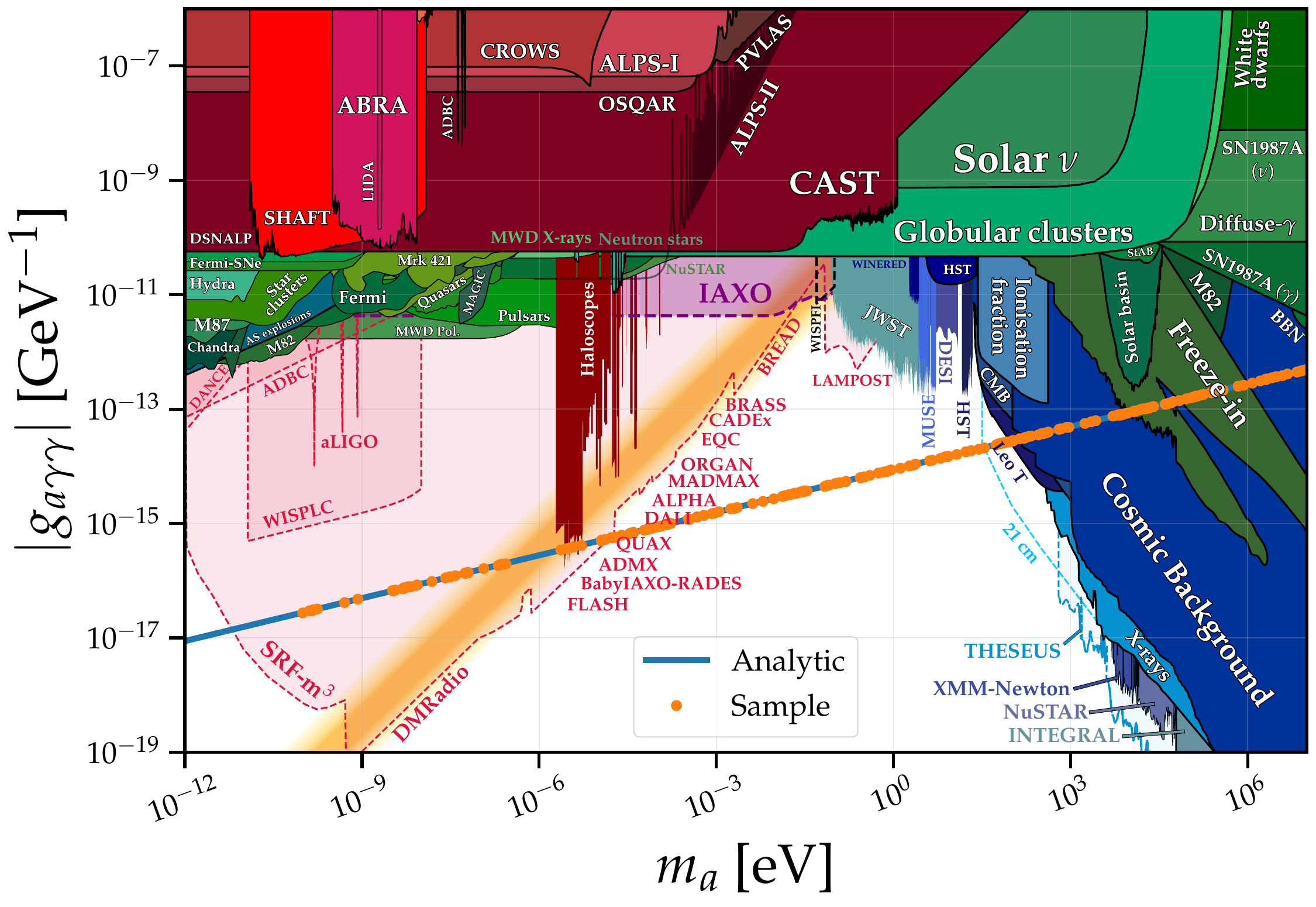}
    \caption{Parameter space of ALPs in the $(m_a,\, g_{a\gamma\gamma})$ plane. 
The solid line shows the analytic relation predicted in Eq.~\eqref{eq:gaggma}, while the dots represent viable points obtained from the parameter scan. 
Shaded regions indicate existing experimental and astrophysical constraints, together with projected sensitivities of future searches \cite{AxionLimits}.}
    \label{fig:axionlike}
\end{figure}

Requiring that the ALP accounts for the entire DM abundance, 
$\Omega_c h^2 = 0.1200 \pm 0.0012$ within the $3\sigma$ range of Planck measurements~\cite{Planck:2018vyg}, 
establishes a direct relation between $m_a$ and $g_{a\gamma\gamma}$. 
Combining Eqs.~\eqref{alpomega} and~\eqref{eq:gagg}, we obtain
\begin{equation}
    g_{a\gamma\gamma} \simeq 
    \frac{g_2^2 \sin^2 \theta_W}{8\pi^2} 
    \frac{N_{\rm DW}\,\theta_a}{10^{16}\,\mathrm{GeV}} 
    \left( \frac{m_a}{4.4 \times 10^{-19}\,\mathrm{eV}} \right)^{1/4},
    \label{eq:gaggma}
\end{equation}
which is shown as the solid blue line in Fig.~\ref{fig:axionlike}, together with existing constraints and projected sensitivities~\cite{AxionLimits}.
Astrophysical observations further restrict the parameter space. 
Current bounds from Leo~T imply $m_a \lesssim 92.6~\mathrm{eV}$~\cite{Wadekar:2021qae}, corresponding to $F_a \gtrsim 4.6 \times 10^{10}~\mathrm{GeV}$. 
Future probes, such as HERA~\cite{Sun:2023acy} and upgraded ADMX~\cite{Graham:2015ouw}, will further constrain the parameter space, leaving an allowed region
$m_a \in [1.8\times10^{-5},\,34.8]~\mathrm{eV}$, 
or equivalently 
$F_a \in [5.9 \times 10^{10},\,2.2 \times 10^{12}]~\mathrm{GeV}$.

Following the same strategy as in the QCD axion case, we perform a parameter scan, and the viable points are shown as orange dots in  Fig.~\ref{fig:axionlike}. 
Imposing perturbativity of $g_2$ up to the Planck scale and the above constraints on $F_a$, we obtain
\begin{equation}
    0.32 (1.12)\times 10^{11}~\mathrm{GeV} \lesssim f_g \lesssim 10^{16}~\mathrm{GeV},
\end{equation}
for current (projected) limits.

A key conclusion is that  
the allowed ranges of $f_g$ (and similarly $f_a$) overlap substantially with those obtained in the QCD axion case. 
Since the GW spectrum from CSs and PTs is primarily determined by these scales, the predicted signals in the two scenarios become effectively degenerate. 
We therefore conclude that GW observations alone cannot distinguish between the QCD axion and generic axion-like realizations within this framework.\\

\noindent \textit{\textbf{Conclusions.}}—
In this work, we have presented a unified analysis of GW signatures in the high-quality axion framework, including contributions from CSs and first-order PTs. 
A key structural feature of this setup is that gauge invariance forbids any explicit bias term capable of lifting the vacuum degeneracy. 
As a result, DWs are generically stable for $N_{\rm DW}\neq 1$, and we are forced to focus on the phenomenologically viable case $N_{\rm DW}=1$, where no GW signal from DW decay is expected.

For the QCD axion, imposing that it accounts for the entire DM abundance and satisfies the high-quality condition leads to a highly constrained symmetry-breaking scale, $f_g \in [1.6\times 10^{11},\,10^{16}]~\mathrm{GeV}$. 
This narrow window directly translates into a well-defined band of GW signals from CSs. 
Remarkably, part of this parameter space yields a signal compatible with current PTA observations. 
In contrast, GWs from PTs are typically peaked at very high frequencies, $f^{\rm peak} \gtrsim 10^7\,\mathrm{Hz}$, and thus probe complementary regions of the model parameter space when combined with different GW experiments.

The situation is qualitatively different for axion-like realizations. 
In high-quality quintessence axion and fuzzy DM axion scenarios, the PQ symmetry is broken before inflation, and all primordial GW signals are exponentially diluted. 
For generic post-inflation ALPs, the $[U(1)_a]\times[SU(2)_{\rm L}]^2$ anomaly induces an axion–photon coupling, allowing laboratory and astrophysical searches to constrain the parameter space. 
Requiring that the axion constitutes all DM implies $\mathcal{O}(10^{11})~\mathrm{GeV} \lesssim f_g \lesssim 10^{16}~\mathrm{GeV}$, 
for both current and projected sensitivities.

A central outcome of our analysis is that the allowed ranges of $f_g$ and $f_a$ in the QCD axion and axion-like cases largely overlap. 
Since the GW spectra from CSs and PTs are primarily controlled by this scale, the resulting signals are highly degenerate. 
We therefore conclude that GW observations alone are insufficient to discriminate between these two classes of axion models, highlighting the necessity of complementary probes.\\

%%%%%%%%%%%%%%%%%%%%%%%%%%%%
\noindent \textit{\textbf{Acknowledgments.}}---The work of R.Z. is supported by the NSFC under Grant No. 12305109, and by Science and Technology Research Project of Chongqing Municipal Education Commission under Grant No.KJQN202300614. The work of J.-W.W. was supported by the NSFC under Grants No. 12405119, the Natural Science Foundation of Sichuan Province under Grant No. 2025ZNSFSC0880, and Fundamental Research Funds for the Central Universities (Grant No. Y030242063002070).
The work of L.B. is supported by the National Natural Science Foundation of China under (NSFC) Grants Nos.12322505, 12547101, L.B. also acknowledges Chongqing Natural Science Foundation under Grant No. CSTB2024NSCQ-JQX0022 and Chongqing Talents: Exceptional Young Talents Project No. cstc2024ycjhbgzxm0020.

\newpage
\bibliography{axionPT}

%%%%%%%%%%%%%%%%%SUPPLEMENTAL MATERIAL%%%%%%%%%%
\newpage
\onecolumngrid
\fontsize{12pt}{14pt}\selectfont
\setlength{\parindent}{15pt}
\setlength{\parskip}{1em}
\newpage
\begin{center}
	\textbf{\large Probing High-Quality Axions with Gravitational Waves} \\ 
	\vspace{0.05in}
	{ \it \large Supplemental Material}\\ 
	\vspace{0.05in}
	{Ruiyu Zhou$^1$, Jin-Wei Wang$^{2, *}$, and Ligong Bian$^{3, \dagger}$} \\
	\vspace{0.05in}
    { \it \small $^1$ School of Science, Chongqing University of Posts and Telecommunications, Chongqing 400065, P.R. China \\
    $^2$ School of Physics, University of Electronic Science and Technology of China, Chengdu 611731, China \\
    $^3$ Department of Physics and Chongqing Key Laboratory for Strongly Coupled Physics, Chongqing University, Chongqing 401331, P. R. China}
    \vspace{0.05in}
\end{center}
\setcounter{page}{1}

In this Supplemental Material, we summarize the explicit form of the one-loop finite-temperature effective potential used in our numerical analysis. All phase transition quantities are computed from the full one-loop effective potential with daisy (ring) resummation in the $\overline{\rm MS}$ renormalization scheme.

The effective potential is given by
\begin{equation}
    V_{\text{eff}}(\phi_1,\phi_2,T)=V_\text{{tree}}(\phi_1,\phi_2)+V_\text{{CW}}(\phi_1,\phi_2)+V_\text{{c.t}}(\phi_1,\phi_2) 
    +V_1^\text{T}(\phi_1,\phi_2,T)+V_1^\text{{daisy}}(\phi_1,\phi_2,T),
\end{equation}
where each contribution is specified below.

\noindent\textit{Tree-level potential.}—
The tree-level potential $V_\text{{tree}}(\phi_1,\phi_2)$ is given in Eq.~\eqref{eq:vp}. Minimizing it yields the vacuum expectation values
\begin{equation}
    v_{\phi_1} = |\langle \phi_1 \rangle| = \frac{f_1}{\sqrt{2}}, \quad v_{\phi_2}=|\langle \phi_2 \rangle| = \frac{f_2}{\sqrt{2}},
\end{equation}
where
\begin{equation}
      f_1=\sqrt{\frac{4\lambda_2 m_1^2-2\kappa m_2^2}{4\lambda_1 \lambda_2-\kappa^2}},  \quad f_2=\sqrt{\frac{4\lambda_1 m_2^2-2\kappa m_1^2}{4\lambda_1 \lambda_2-\kappa^2}}.
      \label{eq:f1f2vev}
\end{equation}

\noindent\textit{Coleman--Weinberg potential.}—
The one-loop zero-temperature correction is given by~\cite{Coleman:1973jx}
\begin{equation}
	V_{\rm CW}(\phi_1,\phi_2)  = 
   \sum_{i} \frac{g_{i}(-1)^{b}}{64\pi^2}  m_{i}^{4}(\phi_1,\phi_2)\left(\mathrm{Log}\left[ \frac{m_{i}^{2}(\phi_1,\phi_2)}{\Lambda_{UV}^2} \right] - C_i\right)\,, \nonumber 
	\label{eq:oneloop}
\end{equation}
where $b=0$ $(1)$ for bosons (fermions), $\Lambda_{UV}$ is the renormalization scale, and $g_i$ denotes the number of degrees of freedom. In our model, $g_i=\{1,1,1,1,3\}$ for $\{\phi_1,\phi_2,G_{\phi_1},G_{\phi_2},Z'\}$, while $C_i=5/6$ for gauge bosons and $C_i=3/2$ otherwise.

\noindent\textit{Counterterms.}—
The counterterm potential $V_\text{{c.t}}(\phi_1,\phi_2)$ is introduced to ensure that the vacuum configuration in Eq.~\eqref{eq:f1f2vev} remains unchanged after radiative corrections:
\begin{align}
V_{\rm c.t}= \delta k \phi_1^2 \phi_2^2 + \delta \lambda_1 \phi_1^4+ \delta \lambda_2 \phi_2^4 + \delta_{m1} \phi_1 ^2 + \delta_{m2} \phi_2 ^2 \;.
\end{align}
The counterterm coefficients are fixed by the renormalization conditions
\begin{align}\label{eq:ct}
    &\frac{\partial V_{\rm c.t}}{\partial \phi_1} +\frac{\partial V_{\rm CW}}{\partial \phi_1} |_{\phi_1 \to v_{\phi_1},\phi_2 \to v_{\phi_2}}=0\;, \nonumber 
    &\frac{\partial V_{\rm c.t}}{\partial \phi_2} +\frac{\partial V_{\rm CW}}{\partial \phi_2} |_{\phi_1 \to v_{\phi_1},\phi_2 \to v_{\phi_2}}=0\;,\nonumber \\
&\frac{\partial^{2} V_{\rm c.t}}{\partial \phi_1 ^2} +\frac{\partial^{2} V_{\rm CW}}{\partial \phi_1 ^2 } |_{\phi_1 \to v_{\phi_1},\phi_2 \to v_{\phi_2}}= 0 \;,\nonumber  &\frac{\partial^{2} V_{\rm c.t}}{\partial \phi_2 ^2} +\frac{\partial^{2} V_{\rm CW}}{\partial \phi_2 ^2 } |_{\phi_1 \to v_{\phi_1},\phi_2 \to v_{\phi_2}}= 0\;, \nonumber \\
&\frac{\partial^{2} V_{\rm c.t}}{\partial \phi_1 \partial \phi_2 } +\frac{\partial^{2} V_{\rm CW}}{\partial \phi_1 \partial \phi_2 } |_{\phi_1 \to v_{\phi_1},\phi_2 \to v_{\phi_2}}= 0.
\end{align}

\noindent\textit{Finite-temperature potential.}—
The one-loop thermal correction is given by~\cite{Dolan:1973qd}
\begin{align}
\label{potVth}
 V_{1}^{T}(\phi_1 ,\phi_2, T) = \frac{T^4}{2\pi^2}\, \sum_i n_i J_{B}\left( \frac{ m_i^2(\phi_1 ,\phi_2)}{T^2}\right)\;,
\end{align}
where
\begin{align}
\label{eq:jfunc}
J_{B}(y) =  \int_0^\infty\, dx\, x^2\, \ln\left[1 - {\rm exp}\left(-\sqrt{x^2+y}\right)\right]\; .
\end{align}

For numerical evaluation, we use the equivalent Bessel-function representation~\cite{Anderson:1991zb}
\begin{align}
\label{Bessel_JFJB}
J_{B}(y) = \lim_{N \to +\infty} - \sum_{l=1}^{N}   \frac{y}{l^2}K_{2} (\sqrt{y} l)\;.
\end{align}

\noindent\textit{Daisy resummation.}—
The daisy (ring) contribution is included as~\cite{Carrington:1991hz,Arnold:1992rz}
\begin{align}
V_{1}^{\rm daisy}(\phi_1 ,\phi_2,T)=&-\frac{T}{12 \pi} \sum_{i={\rm bosons}} n_i \bigg[ ( m_i^2(\phi_1 ,\phi_2)+c_i(T))^{\frac{3}{2}} - ( m_i^2(\phi_1 ,\phi_2))^{\frac{3}{2}} \bigg]\;, \label{DaisyTerms}
\end{align}
where $c_i(T)$ are the thermal masses. In our model, they are given by 
\begin{align}
    c_{\phi_1}(T)&=\frac{1}{12}T^2 (3 q_1^2 g^2+4\lambda_1+k+i\times12y_i) \nonumber\;,\\
    \quad c_{\phi_2}(T)&=\frac{1}{12}T^2 (3 q_2^2 g^2+4\lambda_2+k+j\times12 y^\prime_j) \;,\;\nonumber \\ c_{A_1^\prime}(T)&=\frac{1}{3} q_1^2 g^2 T^2+\frac{1}{3} q_2^2 g^2 T^2\;.
\end{align}

\end{document}